\documentstyle [11pt]{article}

\begin{document}
%
%
\newcommand{\Abs}[1]{|#1|}
\newcommand{\EqRef}[1]{(\ref{eqn:#1})}
\newcommand{\FigRef}[1]{fig.~\ref{fig:#1}}
\newcommand{\Abstract}[1]{\small
   \begin{quote}
      \noindent
      {\bf Abstract - }{#1}
   \end{quote}
    }
\newcommand{\FigCap}[2]{
\ \\
   \noindent
   Figure~#1:~#2
\\
   }
%
%
%
%
\title{On the duality between periodic orbit statistics and quantum level
statistics}
\author{Per Dahlqvist\\
Mechanics Department \\
Royal Institute of Technology, S-100 44 Stockholm, Sweden\\[0.5cm]
}
\date{}
\maketitle

\ \\ \ \\

%
\Abstract{
We discuss consequences of a recent observation that
the sequence of periodic orbits in a chaotic billiard
behaves like a poissonian
stochastic process on small scales.
This enables the semiclassical form factor
$K_{sc}(\tau)$ to agree with predictions of random matrix theories for other
than infinitesimal $\tau$ in the semiclassical limit.
}

\ \\

\section{Introduction}

The spectral form factor $K(\tau)$, defined as the Fourier transform of
the spectral autocorrelation function, has a central
role in semiclassical analysis of chaotic systems.
All spectral statistics bilinear in the density,
such as spectral rigidity ($\Delta_3$) and number variance ($\Sigma_2$)
may be expressed in terms
of the form factor \cite{Brig,Dys}.
The form factor is a convenient tool for semiclassical analysis while
it can be expressed as a double sum over periodic orbits \cite{Brig}.
The approximation of the form factor thus obtained $K_{sc}(\tau)$ is called the
semiclassical form factor.
This paper contains some complementary results to Berry's classical
paper \cite{Brig}, where he shows that, in the limit of the small $\tau$,
the form factor is $K_{sc}(\tau)=\tau$ for system with no time reversal
symmetry and $K_{sc}(\tau)=2\tau$ for system with time reversal symmetry.
This agrees with predictions from random matrix theories \cite{Meht} for the
Gaussian Unitary Ensemble (GUE) and Gaussian Orthogonal Ensemble (GOE)
respectively.
This result is essentially the only semiclassical support for
universality in level statistics for chaotic systems obtained so far.
However, as we will see, the result is only obvious for smaller and smaller
$\tau$ when the semiclassical limit is approached,
and we need a mechanism to ensure validity up to fairly large
$\tau$ (of the order of unity) for universality to be achieved.

In this paper we are going to describe the sequence of periodic orbit
in a statistical language and use tools from the thermodynamic
formalism of chaotic systems \cite{Beck}.
These tools will be worked out in section
2. The main point is assumption B, which states that the sequence of
periodic orbits may be described as a poissonian process, at least on small
scales. In section 3 we show that, under this assumption,
Berry's result $K=(2)\tau$ may be recovered without assuming conditional
convergence of the Gutzwiller formula \cite{Gut}.
And even more important, this result holds for other than infinitesimal
$\tau$ in the semiclassical limit.
We do not
discuss when assumption B breaks down
and the corresponding deviation
from $K=(2)\tau$ may start. In section 4 we discuss the various assumptions
behind our result and how the results depend on them. In particular we
review the numerical evidence for assumption B and
in section 5 we discuss related work
in view of our achievements.

\section{Preliminaries}

The key step is the analytical continuation of
the Gutzwiller formula for the level density down to the real energy axis.
We are going to perform this for the Fourier transform of the level density
under some simplifying assumptions.

\subsection{The Fourier transform of the level density}

Our starting point is
Gutzwiller-Voros
zeta function \cite{Gut,Vor} which for systems with two
degrees of freedom reads
\begin{equation}
     Z(E)=\prod_{p}\prod_{j=0}^{\infty}
          \left(1-\frac{e^{iS_p(E)}}
    {\Abs{\Lambda_{p}}^{1/2} \Lambda_{p}^{j}}\right)
          \ \ ,                       \label{eqn:ZGV}
\end{equation}
where $\Lambda_p$ is the expanding eigenvalue of
the Jacobian along the flow.
To make things as transparent as possible we restrict ourselves to billiards,
so that the action integral is simply $S_p(E)=l_p\cdot \kappa (E)$
where the 'momentum' is $\kappa =\sqrt{2E}$.
The level density to be discussed is measured in
$\kappa$-space and not, as is usual, in
$E$-space.
We use units such that
$m=\hbar =1$.
We have neglected the Maslov indices,
this neglect is discussed in sections 2.2 and 4.

It is expected that the zeros of the zeta function approximate the quantum
eigenvalues. How the the semiclassical approximation and the
stationary phase approximation behind $Z_{GV}$ affect
the spectrum is not all
clear. However, this is not our main concern here.
What is important in this paper is that the zeta function
may contain additional zeros reflecting its convergence properties.

In this paper we will use the following  simplifying assumption:\\
\underline{Assumption A:}\\
{\em The zeta function $Z(\kappa)$
is entire and has zeros on the exact quantum positions $\pm \kappa_i$,
$i\neq 0$
and one extra zero $\kappa_0=ih_{1/2}$ on the positive imaginary
axis lying on the border of
convergence.}
The motivation and discussion of this assumption is postponed
until sec 4. (The reason behind the choice of subscript of $h_{1/2}$ will be
obvious in section 2.2.)

The level density is usually split up into the mean density
and an oscillating part $d(\kappa)=d_0(\kappa)+d_{osc}(\kappa)$.
The leading part of $d_0$ is the Weyl term
$d_0 \sim A\kappa/2\pi $ where $A$ is the billiard area.
The Gutzwiller-Voros zeta function above is derived from the Gutzwiller
formula for $d_{osc}$.
For later purposes we need the Fourier transform of the oscillating part of the
level density
\begin{equation}
\tilde{D}_{osc}(l)=
\frac{1}{2\pi i} \int_{-\infty+iC}^{\infty+iC}e^{-i\kappa l}
\frac{d}{d\kappa} \log Z(\kappa)
d\kappa \ \ .   \label{eqn:Dosc}
\end{equation}
We will evaluate it
in two ways. First, by inserting the product representation of the zeta
function.
It is thus essential the the constant $C$ in eq
\EqRef{Dosc} is sufficiently large so that the contour runs in the region where
the product \EqRef{ZGV} converges (and thus well above all zeros of
$Z$). One can then exchange summation and integration and establish
the result.
\begin{equation}
\tilde{D}_{osc}(l)=
\sum_p l_p \sum_{n=1}^{\infty} \frac{\delta (l-nl_p)}{|M_p^n-I|^{1/2}} \ \ .
\end{equation}
Exponential divergence of this sum $\tilde{D}_{osc}(l) \rightarrow \exp
(h_{1/2}l)$, a feature to be discussed in section 2.2,
is directly related to the presence of the zero $\kappa_0=ih_{1/2}$
as can be seen from eq. \EqRef{Dosc}

Secondly we compute $\tilde{D}_{osc}$ by means of residue calculus
\begin{equation}
\tilde{D}_{osc}(l)=
e^{h_{sc}l}+\sum_{j=-\infty,j\neq0}^{\infty}e^{-i\kappa_j l} \ \ .
\label{eqn:Dosc3}
\end{equation}
The reader may wonder why the contribution from the mean distribution
$d_0$ has disappeared in eq \EqRef{Dosc3}. The reason is that we have neglected
the contribution from the large semi-circle which would have yielded
delta functions, and even more nasty things,
associated with the Fourier transform
of $d_0$.
The tilde-sign in $\tilde{D}_{osc}$ indicates that the result carries
contribution from the extra zero $\kappa_0$. Without this zero we
get
the Fourier transform of the true level density
$D_{osc}=\sum_{j=-\infty,j\neq0}^{\infty}e^{-i\kappa_j l}$.
We can then establish the identity
\begin{equation}
D_{osc}(l)=
\sum_p l_p \sum_{n=1}^{\infty} \frac{\delta (l-nl_p)}{|M_p^n-I|^{1/2}}
-e^{h_{sc}l} \ \ .
\end{equation}
We have all the way assumed that $l>0$.

{}From now on we will tacitly replace every occurrence of $|M_p^n-I|$ with
$|\Lambda_p|$. The errors thus induced are completely negligible in the
limits we are going to explore.
We must now discuss some properties of the periodic orbits
sums.

\subsection{Some properties of the set of periodic orbits}

Essentially all dynamical information is encoded in the sequence
of the invariants of the primitive periodic orbits $\{ l_p,\Lambda_p \}$ (the
maslov indices provide some topological information though).
In the asymptotic limit $l \rightarrow \infty$
one can establish the following family of sum-rules for chaotic systems
\begin{equation}
\sum_p l_p \sum_{n=1}^{\infty} \frac{\delta (l-nl_p)}{|\Lambda_p|^{\beta}}
\rightarrow e^{h_\beta l} \ \ . \label{eqn:sumrules}
\end{equation}

(From now on we will reside in the asymptotic limit of large $l$ and write
equality signs instead of arrows.)
The result applies after appropriate smearing of the delta functions.
The entropy-like quantity $h_\beta$ decreases with increasing $\beta$.
The special case $\beta=1$ was discussed already in \cite{Ozo}; for bound
systems one have $h_1=0$.
$h_0$ is the topological entropy \cite{AAC}.
General $\beta$ are discussed in e.g. \cite{AAC,PDsin,PDLA}.
Usually there are some restrictions on $\beta$. For e.g. the
Sinai billiard eq \EqRef{sumrules} is only valid provided that
$-1<\beta \leq 1$ \cite{PDsin}. In our considerations it suffices if eq.\
\EqRef{sumrules} is valid for $1/2<\beta \leq 1$. We expect this property
to hold for any reasonable system.

It is very useful to describe the set of periodic orbits in a statistical
language. For large $l$ the repetitions of shorter orbits are overwhelmed by
the number of primitive orbits so that we may neglect the sum over $n$ above.
Let us call the density of
of prime orbits $\phi(l)=\sum_p \delta (l-l_p)$. From eq.\ \EqRef{sumrules}
we see that the mean value of this density is
\begin{equation}
<\phi (l)>= e^{h_0 l}/l \ \ .
\end{equation}
For general $\beta$ we express the sum rules in terms of the averages
$<|\Lambda_p|^{-\beta}>$
\begin{equation}
l<\phi><|\Lambda|^{-\beta}>=e^{h_0 l}<|\Lambda|^{-\beta}>=e^{h_\beta l}  \  \ .
\end{equation}
For instance we get the result which will be of use later
 \begin{equation}
<|\Lambda|^{-1}>=e^{-h_0 l} \ \ .
\end{equation}
So much for the large scale structure  of the sequence
$\{l_p,\Lambda_p\}$, i.e. large smearing widths
in eq \EqRef{sumrules}. What about the small scale structure?

Let us order this sequence according to increasing $l_p$ and consider
the ordered sequence $\{l_i,\Lambda_i\}$ where the integer $i$ denote the
position in the sequence. Then rescale the length variable
according to
\begin{equation}
\ell_i=\int_0^{l_i} <\phi (l')>dl'   \ \ ,
\end{equation}
so that the mean spacing $<\ell_i -\ell_{i-1}>$ is unity.
We will now make our main assumption.\\
\underline{ Assumption B:}\\
1/ {\em The sequence $\ell_i$ is given by a poissonian process
with unit intensity.}\\
2/ {\em The corresponding stabilities $\Lambda_i$ may be considered
as mutually independent stochastic variables}
We discuss the evidence for this assumption in sec 4.


We can now reformulate $D_{osc}$ in a purely statistical language
\begin{equation}
D_{osc}(l)=l\left( \phi(l)\frac{1}{\sqrt{|\Lambda(l)|}}-
<\phi><\frac{1}{\sqrt{|\Lambda|}}>\right) \ \ . \label{eqn:stat}
\end{equation}

The introduction of phase indices (Maslov indices and symmetry indices
\cite{CEsymm})
will generally move down the leading
zero $h_{1/2}$ \cite{PDLA} and there is a possibility that it might even cross
the real $\kappa$-axis making the Gutzwiller sum conditionally
convergent \cite{Stein}.
It is nontrivial to deduce if this really takes place for a given system.
Many estimations of the
position of the {\em entropy barrier}, with or without
Maslovs, in the literature
assumes uniform hyperbolicity of the system and are thus
invalid for generic systems.

\section{The form factor}

The spectral form factor is defined as the Fourier transform of the
spectral autocorrelation function
\begin{equation}
K=\frac{1}{d_0}\int_{-\infty}^{\infty} d\epsilon e^{-i\epsilon l}
d_{osc}(\kappa+\epsilon/2)d_{osc}(\kappa-\epsilon/2)   \ \ .
\end{equation}
It is usually
regarded as a function of the
dimensionless length variable
$\tau=l/(2\pi d_0)=l/(A\kappa)$
with $\kappa$ as a parameter. The suggested universal
behaviour of $K(\tau )$ should arise in the semiclassical limit
$\kappa \rightarrow \infty$.
To be meaningful the form factor needs some averaging
which we will apply
first at the end.
The form factor can be expressed in terms of the Fourier
transform of $d_{osc}(\kappa)$ according to
\begin{equation}
K=\frac{1}{2\pi d_0} \int_{-\infty}^{\infty} dl'
\int_{-\infty}^{\infty} dl'' \delta(l-\frac{l'+l''}{2})
cos(\kappa(l'-l'')) D_{osc}(l')D_{osc}(l'')  \ \ . \label{eqn:K}
\end{equation}
The derivation is straightforward, one has to use the fact that
$d_{osc}(\kappa)$, and thus $D_{osc}(l)$, are real and even.

\subsection*{Assuming conditional convergence}

Let us now follow Berry's arguments a little longer \cite{Brig}.
If the Gutzwiller formula is conditionally
convergent we can insert $\tilde{D}_{osc}$ directly instead of
$D_{osc}$ into \EqRef{K}.
\begin{equation}
K_{sc}=\frac{l^2}{2\pi d_0} \int \int dl' dl'' cos(\kappa(l'-l''))
\delta(l-\frac{l'+l''}{2})
(\sum_i \frac{\delta(l'-l_i)}{\sqrt{|\Lambda_i|}})
(\sum_j \frac{\delta(l''-l_j)}{\sqrt{|\Lambda_j|}}) \ \ . \label{eqn:Kcond}
\end{equation}
Keeping $\kappa$ constant and letting $l\rightarrow 0$ be small the
cosine will wash away the non diagonal terms (assuming no systematic
degeneracies among the $l_i$ due to e.g. time reversal symmetry)
and we get
\begin{equation}
K_{sc}=\frac{l^2}{2\pi d_0}
(\sum_i \frac{\delta(l-l_i)}{|\Lambda_i|}) \ \ .
\end{equation}
Using the sum rules in sec 2.2 we get
\begin{equation}
\bar{K}=\frac{l^2}{2 \pi d_0 } \frac{1}{l}
\exp(-h_0 l)=\frac{l}{2\pi d_0} = \tau \ \ . \label{eqn:KGUE}
\end{equation}

This result gave rise to some enthusiasm since it agrees with predictions of
random
matrix theories.
But one may ask two questions.
First, is eq \EqRef{KGUE} true even if
the Gutzillwer sum is not conditionally convergent?

Secondly, we note that
the number of periodic orbits contained in
one period of the cosine in eq \EqRef{Kcond}
is
$\sim <\phi>/{\kappa}\sim \exp(h_0A\kappa \tau)/(A\kappa^2 \tau)$.
To make this estimate we have assumed that the smearing width is
$\Delta l \approx <\phi>^{-1}$
which is the smallest conceivable choice.
In the semiclassical limit ($\kappa\rightarrow \infty$)
the argument leading to eq \EqRef{KGUE} is brutally violated
for other than infinitesimal $\tau$.
If the predictions of random matrix theories are correct one
expects eq \EqRef{KGUE} to hold up to
$\tau \sim 1$. Now to the second question. Is
the result $K=\tau$ correct
for finite $\tau$, and if the answer is yes, why?

\subsection*{The generic case}

The key lies in the stochastic nature of the periodic orbits as formulated
in assumption B. First we insert the general formula for
$D_{osc}(l)$
\begin{equation}
\begin{array}{ll}
K_{sc}=& \frac{l^2}{2\pi d_0} \int \int dl' dl'' cos(\kappa(l'-l''))
\delta(l-\frac{l'+l''}{2}) \cdot \\
&
(\sum_i \frac{\delta(l'-l_i)}{\sqrt{|\Lambda_i|}}-
<\phi><\frac{1}{\sqrt{|\Lambda|}}>)
(\sum_j \frac{\delta(l''-l_j)}{\sqrt{|\Lambda_j|}}-
<\phi><\frac{1}{\sqrt{|\Lambda|}}>) \ \ .
\end{array}
\end{equation}
We clearly see that we are dealing with correlation in the
sequence $\{l_i,\Lambda_i\}$ and, according to assumption B, there are no
correlations at all.
Therefore only the diagonal terms will contribute.
In order to avoid delta functions in the resulting form factor
we smear it
\begin{equation}
\bar{K}_{sc}=\frac{1}{\Delta}\int_l^{l+\Delta}K(l')dl'  \ \ ,
\end{equation}
which now may be expressed as
\begin{equation}
\bar{K}_{sc}=\frac{l^2}{2 \pi d_0 \Delta}\int_l^{l+\Delta}dl'
\left\{
(\sum_i \frac{\delta(l'-l_i)}{\sqrt{|\Lambda_i|}}-
<\phi><\frac{1}{\sqrt{|\Lambda|}}>)  \right\}^2  \ \ .
\end{equation}
The integral in this expression is just
the variance of the sum of
$1/\sqrt{|\Lambda|}$ in a window of a poissonian process,
\begin{equation}
\bar{K}=\frac{l^2}{2 \pi d_0 \Delta}V_\Delta(\sum \frac{1}{\sqrt{|\Lambda|}}) \
\ .
\end{equation}
The calculation of this variance is an elementary exercise in probability
theory (see Appendix) and the result is
$V_\Delta =\lambda \Delta <1/\sqrt{|\Lambda|}^2>$.
The intensity $\lambda$ equals the mean density of prime orbits
$\lambda = <\phi(l)> =\exp(h_0 l)/l$ and according to the sum rules in sec
2.2 we have\\
$<1/\sqrt{|\Lambda|}^2>=<1/\Lambda>=\exp(-h_0 l)$.
We thus get our final result
\begin{equation}
\bar{K}_{sc}=\frac{l^2}{2 \pi d_0 \Delta} \frac{\exp(h_0 l)\Delta}{l}
\exp(-h_0 l)=\frac{l}{2\pi d_0} = \tau  \  \ ,
\end{equation}
which is the same as for conditionally convergent systems
but the result now holds for other than infinitesimal $\tau$ in the
semiclassical limit.

It is straightforward to generalize to the
systematic degeneracies of periodic orbit exhibited by time
reversible systems, and we will not discuss it.

\section{Motivations for our basic assumptions}

Our result relies on a series of assumptions and approximations.
Some of them may be removed or modified without altering the result.
In this section we
motivate our assumptions
and discuss the extent to which the results depend on them.

First assumption A. The presence of a leading zero $\kappa_0=ih_{1/2}$
such that
$h_{1/2}>0$ (no Maslovs) has already been discussed.
In the general case it is reasonable to assume the
presence of several zeros not associated with
any quantum state. Their contribution is naturally included into
$<\phi><|\Lambda|^{-1/2}(l)>$ giving rise to oscillatory and
exponentially decreasing corrections.

The assumption that the semiclassical zeros equals the quantum eigenvalues
is not crucial for our results. It was mainly introduced for computational
and notational convenience. However,
the expected failure of the semiclassical eigenvalues to be real
will have consequences
for the large $\tau$ behaviour of
the semiclassical form factor \cite{Keat}, see section 5!

There is one example for which assumption A is exactly fulfilled and that
is compact billiards on surfaces of constant negative curvature.
These systems
are special having
zeros on the exact quantum positions. But there is
also a zero on the imaginary $\kappa$ axis, right on the border of
convergence \cite{DUKE}.
Indeed there is a zero on
the border of convergence of each $j$-factor in the zeta function
\EqRef{ZGV}.

Now to assumption B.
In ref \cite{Tak} the authors pursued the original idea to consider the
spectrum of lengths $l_j$ of the prime cycles and do level statistics
\`{a} la quantum chaos. Their system was a touching three-disk billiard.
The first step is to unfold the spectrum, cf section 2.2, yielding
the sequence $\ell _j$. Then they studied the level spacing distribution, which
was found to be an exponential to very high accuracy, and spectral rigidity,
which was found to agree with $\Delta_3(L)=L/15$. This is consistent with
the sequence $\ell_j$ being given by a poissonian process and motivates
assumption B.
The three-disk billiard, as well as any generic Euclidean chaotic billiard,
has almost certainly an infinite symbolic dynamics. It would be nice to know
if the greater regularity among the cycles for a
finite symbolic dynamics still exhibits this kind of randomness.

We do not attempt to explain this random feature, we
only offer the following hand-waving argument.
Neighbours in the sequence $\{l_j\}$ need not be close in phase space
and there is therefore no reason for correlations.
In a small proximity to a given length $l$
there may be many periodic orbits.
The sequence of periodic orbit could thus, perhaps, be viewed
upon as the random superposition of many sequences, each sequence
corresponding to a topologically distinct family of periodic orbits. This
would give rise to the poissonian nature.

Assumption B2, concerning the the stabilities, is in the same
spirit as B1. If the lengths are uncorrelated so should the eigenvalues be.

The restriction to billiards is mostly for convenience.
We find it highly unlikely that a smooth potential would exhibit
complete chaos. Our calculations would be easily modified
for a chaotic smooth potential, if existing, which is homogeneous;
the 'momentum' variable $\kappa$ being
some other power of energy $E$.
If one chooses to consider non homogeneous potentials one must perform the
Fourier transform with respect to $\hbar$ instead of $\kappa$
We won't speculate about this case since
it would take us to far from the case where assumption B has been verified.

\section{Discussion}

It is interesting to note that the semiclassical form factor $K_{sc}(\tau)$
in the region $0<\tau < 1$ depend on {\em both} the very small scale
structure {\em and} the very coarse structure on the sequence of periodic
orbits.
In this paper we have focused on the deep asymptotic limit $\kappa \rightarrow
\infty$. For finite energies one may have to consider
pre-asymptotic behaviour and power law corrections of the
periodic orbit sum rules involved. This is discussed in
refs \cite{PDreson,PDLA}. In these papers we did not correct for the divergence
of the trace formula
but this procedure is readily justified from the present results.
This pre-asymptotic behaviour extend considerably the non universal regime
in spectral statistics derivable form the form factor as compared to the
regime discussed by Berry. Such non-universal regimes have been observed in
several numerical experiments \cite{Diet,Arve,Mart,PDLA}.
In ref \cite{PDLA} we also discuss the role of marginally stable orbits
which plays a major role for moderate energies.

The semiclassical status of the large $\tau$ limit is much more obscure.
It is clear that
the quantum form factor approaches unity in this limit $K \rightarrow
1$ provided that there are no systematic degeneracies in the spectrum.
In \cite{Keat} Keating demonstrates that,
since we cannot expect the trace formula to produce poles exactly on the
real axis, the semiclassical
form factor should diverge exponentially. If the imaginary part
of the poles is much less than the mean spacing, as is indicated by \cite{TW},
this exponential
take off should not occur until fairly large $\tau$ and one can still
hope for saturation of the semiclassical form-factor.

Some evidence of saturation is presented in ref. \cite{Argam} for
the hyperbola billiard and other systems.

The exponential collapse of
$K_{sc}(\tau)$ reported in \cite{AS} appears to be due
to neglect of the
extra zero(s) and illustrates the hazard of inserting diverging series into the
form-factor. This is particularly dangerous since the expression thus obtained
need not be divergent.

Whatever happens to the form factor we can of course not expect assumption
B to hold throughout the spectrum and a deviation form $K_{sc}(\tau)=(2)\tau$
should be expected. If the expected saturation indeed takes place it is, of
course,
highly desirably to understand its classical origin and its
manifestation by the periodic
orbits and the connection to the fact that the system is bound.

The reader may think that our assumption B contradicts the concept of
{\em action repulsion} discussed by Argaman et.al. \cite{Argam}. However, if
the  periodic orbit correlations proposed in ref. \cite{Argam} really occur
this repulsion is not an effect acting among neighbours in the sequence
$\{ l_i \}$
of cycles but over vast distances, the name action repulsion is thus not very
appropriate.

Much of the present work on the semiclassical trace formula is concerned with
taming the diverging trace formula and
the computation of corrections in order to obtain accurate results for the
bottom part of the spectrum.
Many of these corrections disappear in the semiclassical limit.
In this paper we took the opposite point of view.
We tried to relate asymptotic (=semiclassical) properties of the
spectrum to the asymptotic behaviour of the periodic orbits. As both are
conveniently expressed in a statistical language this approach aims at
unveiling the duality between periodic orbit statistics and level statistics.

\section*{Acknowledgements}

This work was supported by the Swedish Natural Science
Research Council (NFR) under contract no. F-FU 06420-303.

\newcommand{\PR}[1]{{Phys.\ Rep.}\/ {\bf #1}}
\newcommand{\PRL}[1]{{Phys.\ Rev.\ Lett.}\/ {\bf #1}}
\newcommand{\PRA}[1]{{Phys.\ Rev.\ A}\/ {\bf #1}}
\newcommand{\PRD}[1]{{Phys.\ Rev.\ D}\/ {\bf #1}}
\newcommand{\PRE}[1]{{Phys.\ Rev.\ E}\/ {\bf #1}}
\newcommand{\JPA}[1]{{J.\ Phys.\ A}\/ {\bf #1}}
\newcommand{\JPB}[1]{{J.\ Phys.\ B}\/ {\bf #1}}
\newcommand{\JCP}[1]{{J.\ Chem.\ Phys.}\/ {\bf #1}}
\newcommand{\JPC}[1]{{J.\ Phys.\ Chem.}\/ {\bf #1}}
\newcommand{\JMP}[1]{{J.\ Math.\ Phys.}\/ {\bf #1}}
\newcommand{\JSP}[1]{{J.\ Stat..\ Phys.}\/ {\bf #1}}
\newcommand{\AP}[1]{{Ann.\ Phys.}\/ {\bf #1}}
\newcommand{\PLB}[1]{{Phys.\ Lett.\ B}\/ {\bf #1}}
\newcommand{\PLA}[1]{{Phys.\ Lett.\ A}\/ {\bf #1}}
\newcommand{\PD}[1]{{Physica D}\/ {\bf #1}}
\newcommand{\NPB}[1]{{Nucl.\ Phys.\ B}\/ {\bf #1}}
\newcommand{\INCB}[1]{{Il Nuov.\ Cim.\ B}\/ {\bf #1}}
\newcommand{\JETP}[1]{{Sov.\ Phys.\ JETP}\/ {\bf #1}}
\newcommand{\JETPL}[1]{{JETP Lett.\ }\/ {\bf #1}}
\newcommand{\RMS}[1]{{Russ.\ Math.\ Surv.}\/ {\bf #1}}
\newcommand{\USSR}[1]{{Math.\ USSR.\ Sb.}\/ {\bf #1}}
\newcommand{\PST}[1]{{Phys.\ Scripta T}\/ {\bf #1}}
\newcommand{\CM}[1]{{Cont.\ Math.}\/ {\bf #1}}
\newcommand{\JMPA}[1]{{J.\ Math.\ Pure Appl.}\/ {\bf #1}}
\newcommand{\CMP}[1]{{Comm.\ Math.\ Phys.}\/ {\bf #1}}
\newcommand{\PRS}[1]{{Proc.\ R.\ Soc. Lond.\ A}\/ {\bf #1}}

\section*{Appendix}

\AA ke Svensson owns a small shop
in the old town of Stockholm.
Customers arrive at the shop according to a poissonian process
with intensity $\lambda$.
The
amount of money paid by one customer is considered as a
stochastic variable $x$ with
probability distribution $f(x)$. $x$ belonging to different customers
are mutually independent

During time T the  amount of cash received by \AA ke is
$X_T$. What is
the variance of $X_T$?

The distribution of arrivals during time $T$ in a poissonian process is
\begin{equation}
p_n=e^{-\lambda T} \frac{(\lambda T)^n}{n!}  \ \ ,
\end{equation}
so the distribution of the variable $X$ is
\begin{equation}
F(X)=\sum_{n=0}^\infty p_n f^{*n}(X)  \ \ ,
\end{equation}
where $f^{*n}$ is the n-fold convolution of $f$.
The mean and variance in such a convolution are additive so we have
\begin{eqnarray}
<x_n>=n<x> \\
<x_n^2>=n<x^2>+n(n-1)<x>^2  \ \ .
\end{eqnarray}
A short calculation now yields the mean and variance of
$X_T$
\begin{eqnarray}
<X_T>=\lambda T \cdot <x>\\
V(X_T)\equiv <X_T^2>-<X_T>^2=\lambda T \cdot <x^2>  \ \ .
\end{eqnarray}

\end{document}